# Excitation of helical shape argon atmospheric pressure plasma jet using RF pulse modulation

Mahreen Khan, Veda Prakash G., Satyananda Kar*, Debaprasad Sahu and Ashish Ganguli

Department of Energy Science and Engineering, Indian Institute of Technology Delhi, Hauz Khas, New Delhi – 110016, India

*satyananda@ces.iitd.ac.in

**Abstract**

The article reports the excitation of a helical argon atmospheric pressure plasma jet using a pulse modulated 13.56 MHz radiofrequency (RF) power source. This helical structure is observed in open ambient air which is far different from the conventional conical shape. This helical structure originates due to the periodic pressure variation in the discharge region caused by pulse modulated RF (2 kHz modulation frequency ($f_p$)) and propagates downstream into the ambient air. The geometrical characteristics of the observed structure are explored using optical imaging. Moreover, the influence of various input parameters viz., duty cycle (D), gas flow rate (Q), and RF power (P) of the modulated pulse on the formation of helical structure are studied. These helical structures have an implication on the plasma jet chemical features (enhancement of reactive oxygen and nitrogen species (RONS)) as these are involved in increase in air entrainment into the ionization region desired for various plasma applications.

## I. INTRODUCTION

In recent years, cold atmospheric pressure plasma research is being undertaken extensively, as it demonstrates features desirable for biomedical and industrial applications.[1–7] In cold plasma jets, the applied electric field between the two electrodes ignites the plasma which is blown out into the ambient air with an optimum gas flow rate, to form a plasma jet. Most of the plasma jets reported in the literature are in a conical shape to the naked eye.[3,8–10] Under certain experimental conditions, the applied voltage characteristics and gas flow rate yield electro-hydrodynamic features and have been shown to impact the plasma jet shape.[11–22] Despite experimental and theoretical difficulties, understanding these phenomena can offer fundamental insight into dynamics involved with gas flow, excitation source characteristics, and plasma generation. Moreover, a dynamically controlled shape of plasma jets can increase the rate of plasma-chemical features due to ambient gas mixing which has obvious advantages for various plasma applications.[23,24]

With increasing research on cold atmospheric pressure plasmas, distinctive plasma discharge features in the inter-electrode region as well as in the plasma jet tip have been reported. Previously, wave-like and random structures have been observed at the tip of the plasma jet, which are related to the increase in gas velocity caused by the onset of plasma, resulting in a turbulent flow.[11,18,21,22,25] Also, in a pulsed DC helium jet, Liu et. al, have excited helix-shaped plasmas inside the extended dielectric tube by using an external helical coil.[16] They ascertained this wall-hugging plasma propagation to be the effect of the applied voltage. Moreover, stationary and rotating structures in the inter-electrode discharge region in the glass tube have been observed due to thermal interference for RF and microwave argon plasmas.[12,26,27] In argon plasma jet with low-frequency bias, the presence of solid or hollow swellings is reported which depend on the applied voltage polarity where the formation of swells is attributed to discharge enhancement by active species which are provided by intense discharges near the nozzle.[28] Therefore, for pulsed or low-frequency excitation sources, most of the complex structures and plasma dynamics are observed *inside* the discharge tube *or at the plasma jet tip*.

In this paper, we report the generation of helical argon plasma jets without using any external helical coils, rotating electric fields, or external magnetic fields and which, can be controlled through basic input parameters, like the gas flow, RF duty cycle, and power, etc. The helical shape of the argon plasma jet outside the discharge tube is experimentally observed in our pulse-modulated RF plasma jet that originates at the glass nozzle and propagates downstream freely into the atmosphere. Importantly, this plasma jet shape is totally different from those reported earlier, which are usually of conical shape in majority of the cases. The influence of the different input parameters on these structures is experimentally investigated in detail, with the aim of gaining insight into the formation and geometrical characteristics of the helical plasma jets. In addition, this helical structure can also promote enhancement of plasma-chemical features by efficient mixing of the air into the ionized plasma region,[23,24] which is suitable for various plasma applications. Also understanding the phenomenon of helical structure dynamics has an impact in diverse fields which includes, lasers, communications [29], solar atmosphere [30], turbulent flows, biophysics, and other areas.[2,31–33]

The present manuscript is arranged as follows. Section II presents the experimental setup and diagnostics used, Section III discusses the experimental results and discussions, and section IV conclude the results.

## II. EXPERIMENTAL SETUP

Figure 1 shows the schematic diagram of the experimental setup. RF plasma was generated in a pyrex glass tube of 3 mm inner diameter, 1 mm wall thickness, and 10 cm length. A copper powered electrode (dia. ~ 1.5 mm) was introduced into the glass tube (up to 3 mm above the nozzle). A copper strip, wound around the tube just above the nozzle, was used as the ground electrode. Electrodes were arranged in a cross-field configuration. A 13.56 MHz RF power generator (CESAR 1310) was operated in continuous (CW) as well as pulsed mode (PM) (at repetitive pulsing frequency, $f_p$, of 2 kHz, and duty cycle (10- 80 %) to supply RF power (20- 50 W) through a matching network to the plasma reactor. Argon gas was used as the discharge gas with a flow rate from 0.5 - 12 lpm. Applying RF power to the central, RF electrode along with argon gas feed to the plasma reactor, ignites the discharge, which appears as a plasma jet in the ambient air, for both CW and PM.

The applied voltage was measured by using a P6015A Tektronix high voltage probe connected in-between the RF power generator and matching network. A Phantom VEO 410 high-frame-rate (5200 fps) camera was equipped with a Nikon 24-85 mm focal lens to image (two dimensional (2D)) the plasma jet. All images were recorded with 2200 μs exposure time. In this article, the camera was arranged in two different positions: (i) perpendicular to the axial direction to capture the front view of plasma jet (viewing the plasma jet normally), (ii) inclined at a 45° angle to the axial direction facing the discharge (i.e., - z-direction) to capture the helical twist of the plasma jet. Gas temperature was measured using an insulated K-type thermocouple. Acoustic measurements were recorded using a microphone arranged in perpendicular to the plasma jet at a 5 mm distance from the nozzle. The recorded information is processed with the soundcard of a computer and analyzed in the frequency domain using a developed MATLAB code.

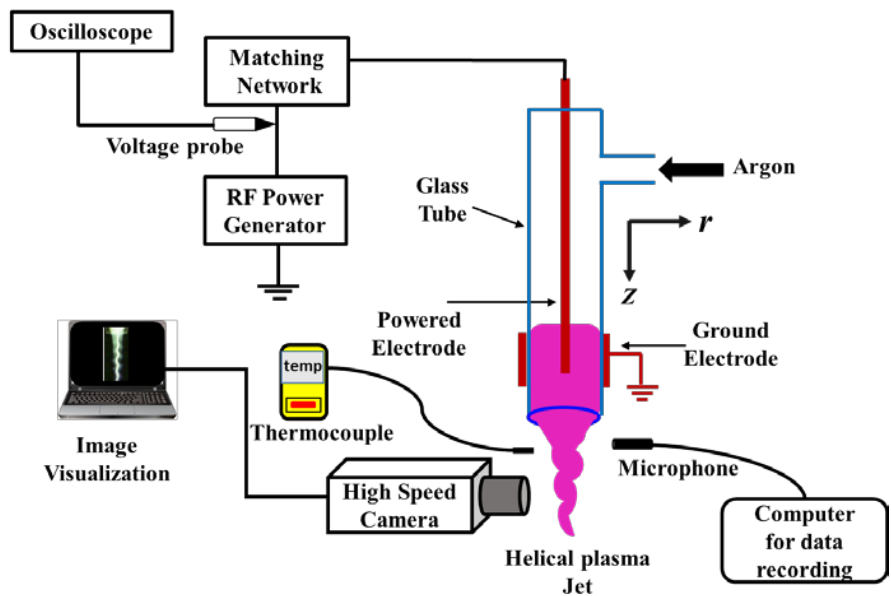

**Figure 1:** Schematic of the experimental setup and diagnostics arrangement

## III. RESULTS AND DISCUSSION

### A. Generation of helical plasma jet

A comparison of the geometrical structures of the RF argon plasma jet in CW and pulsed modes (2 kHz ($f_p$), 30% duty cycle (D)) is shown in Figure 2. In CW mode, (shown in Figure 2(a)), the plasma jet is visible as a smooth and regular conical shape, where the bright central filament surrounded by the glow region (a common feature of argon plasma jets [34,35]) is aligned along the axial direction. On the other hand, for the pulsed mode, a helical-shaped plasma jet is observed which is visible to the naked eye as shown in Figure 2(b). Here the propagation of the central filament surrounded by the glow region collectively takes a helical path around its axis. The clear images of the helical structures were recorded at 2200 μs exposure time for which both filament and glow region are prominently visible. For lower exposure times, viz., 50, 100, 500, 1000 μs (images not shown here), only the filament is visible and the visibility of the glow region is poor. Nonetheless, the helical structure of the filaments is present and seen clearly even for short exposure times. Moreover, the helical structures are also noticed for smaller and larger diameter pyrex tubes with reference to the tube diameter that was used for the experiments.

To figure out the direction from a geometrical point of view, a small part of the clockwise helix can be described with a positive slope, $tan\ \theta > 0$,[36] where $\theta$ is the corresponding angle between the initial side (p) and the terminal side (q) as shown in Figure 2(b). In the present case, $\theta$ corresponds to ~ 45° which gives the slope of '1'. Thus this indicates the helical structure observed has a clock-wise direction. The helical pattern can be seen in greater clarity in the inset of Figure 2 (b), captured from a 45˚ viewing angle as described in Section II above. It may be noted that the helical phenomenon reported here is *from the nozzle downwards*, rather than *inside the glass tube* [16,26] or *near the plasma jet tip* [11] as reported in the literature. The direction of the helical path with reference to its axis is acquired immediately after the discharge and remains same for the entire experiment. This shape is repeatable and the structure shape remains unaltered until either of the inputs i.e., RF power and gas supply is varied or turned off. Furthermore, it is emphasized that a significant amount of acoustic emission is noticed in the audible range with the excitation of these helical structures.

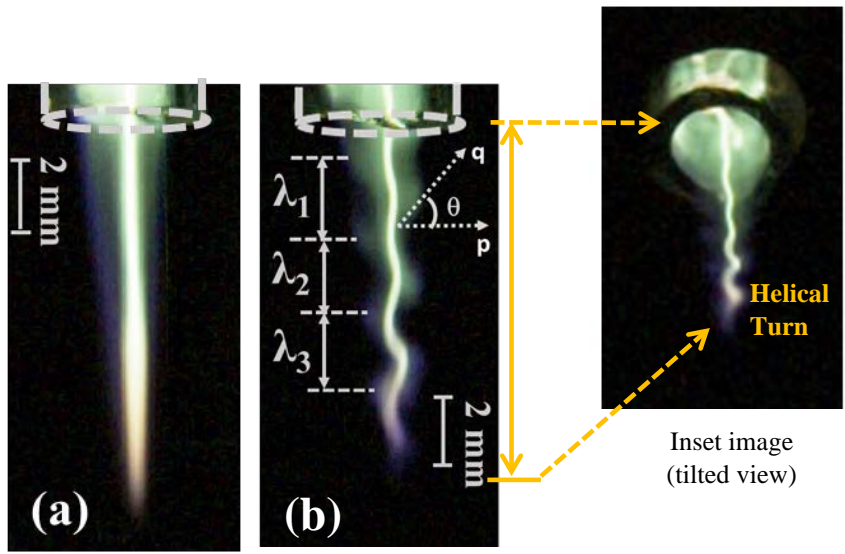

**Figure 2:** Typical 2D front view images of a plasma jet in ambient air. (a) CW RF mode (P= 30 W, Q= 1.5 lpm, plasma length ~ 12 mm). (b) PM RF mode (P= 50 W, $f_p$= 2 kHz, D= 30%, Q=1.5 lpm. Pitch ($\lambda_1, \lambda_2, \lambda_3$ ~ 2.3 mm) and length ~ 10 mm). The image was captured by arranging the camera at perpendicular to the axial direction to capture the front view. In the image, the positive slope $tan\ \theta > 0$ indicates clockwise direction of helix. In the present case, $\theta$ ~ 45°. The inset image shows a tilted image from the bottom captured by pointing the camera ~ 45˚ angle to the axis, facing the discharge (i.e., -*z*-direction in Figure 1). It clearly shows the filament propagating in a clockwise helical path, surrounded by a glow discharge.

**B. Geometrical features of the helical plasma jet:**

The geometrical features and propagation of the helical plasma jet are studied by analyzing the images using Image J (1.52 a version) software. As shown in Figure 3(a), at 50 W applied power the generated plasma jet has 4 helical turns with 2.3 mm pitch (λ) between adjacent turns and remains almost constant over the plasma jet length. Viewing the jet from the left in the figure, the letters A, C, E, G indicate bulging or bending of the jet towards the left, hitherto called the 'bend' regions, while the letters, B, D, F indicate bulging in the opposite direction (towards the right), referred hitherto, as the 'gap' regions (when viewed from the left). The corresponding normalized intensity variation (with the radial distance) is shown in Figure 3 (b). The locations of the bends (shaded red), gaps (blue), and the filament peak are also shown in Figure 3(b). Image analysis clearly shows a greater radial extension of the image at the bulges in comparison to that at the gaps (where the image expands more towards the opposite side). In the figure, *the clockwise rotation of the bulge* as the jet propagates downwards, is illustrated using the crosses '⊗' and the dots '⊙'. This recurrence of the radial expansion of the image towards alternate sides with a constant pitch along the axis and the clockwise rotation of the bulge offers clear evidence of propagation of the plasma jet along a helical path in the present experiments.

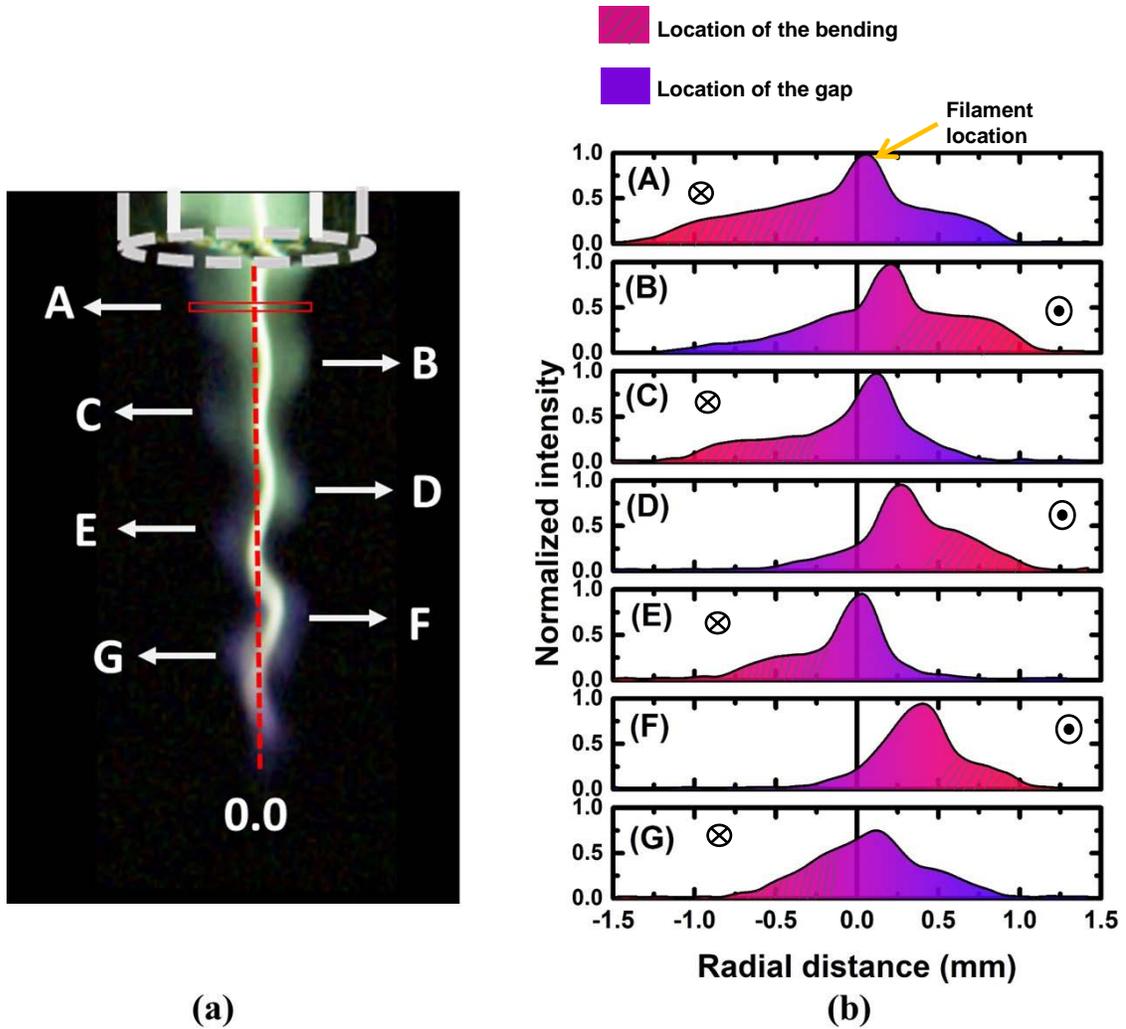

(a)  (b)

**Figure 3**: (a) shows a front view image of the location of the plasma bending ('A'-'G') due to helical turns in plasma jet. The vertical dotted line and horizontal bar represent the reference axis

("0.0" location) and the region of interest (ROI) respectively (b) The spatial intensity (for ROI) of the plasma jet shows the alternate variations in intensity from locations 'A'– 'G' with reference to the axis of the plasma jet. To indicate the clockwise rotation of the helical jet, the symbols cross '⊗' and dot '⊙' have been used to denote the direction of the curling of the fingers (with thumb pointing down vertically) 'into' and 'out' of the plane of the paper.

### C.     Mechanism of helical plasma jet formation

A possible explanation for the helical plasma jet formation in the pulsed mode may be given as follows. The formation of the helical plasma jet (and consequently the production of the acoustic pulse as well) could be attributed to the combined effect of periodic localized heating and cooling of the neutral gas during plasma on and off durations respectively. Gas heating majorly is generated in atmospheric pressure plasmas through elastic collisions between electron and neutral gas particles (Ohmic heating), where highly energetic electrons transfer energy to neutral atoms, resulting in the increase of the neutral gas temperature.[37–39] For the pulsed mode used in the present experiments, the total period of the applied pulses is 500 µs (for 2 kHz pulsing frequency), so that for a 30% duty cycle, the applied RF power remains on for 150 µs ($T_{on}$) and off for 350 µs ($T_{off}$). Typical pulse modulated voltage waveform depicting the same is shown in Figure 4. The repetitive RF pulses result in the formation of pulsed plasma or plasma pulses, that lead to alternate sequences of gas heating (plasma expansion) and cooling (plasma contraction) in the on and off periods respectively. In the 'on' duration of the RF pulse, electron production around the powered electrode (by electron impact ionization) increases, resulting in enhanced depletion and localized heating of the neutral gas in the annular discharge region. Meanwhile, in the 'off' duration, the plasma quenches and the gas cools down. As a result, the neutral gas fluid experiences a corresponding variation of gas temperature and gas density within the discharge region. This leads to the excitation of a periodic pressure gradient between the central core (on the $z$-axis) and the radial direction (along the $r$-axis). A natural consequence of these *pulsed pressure gradients* above a certain threshold, is the production of audible acoustic signals.[40] In the present experiments also a significant amount of acoustic emission in the audible range was observed from the plasma. To understand the features of the acoustic, it was recorded using a microphone and analyzed in the frequency domain with the developed MATLAB code. The frequency spectrum of the acoustic measurements shows the presence of 2 kHz and its higher-order harmonics as shown in Figure 5. It is expected that these originated from temporal variations of the local pressure in the discharge region. Generation of such acoustic signals in atmospheric pressure plasma jets due to the pulsed excitations are also reported in the literature.[41–43] Recently Platier et al.[43] were able to experimentally measure this pressure variation related to the acoustic waves in pulsed RF driven plasma jet, which further supports the present context.

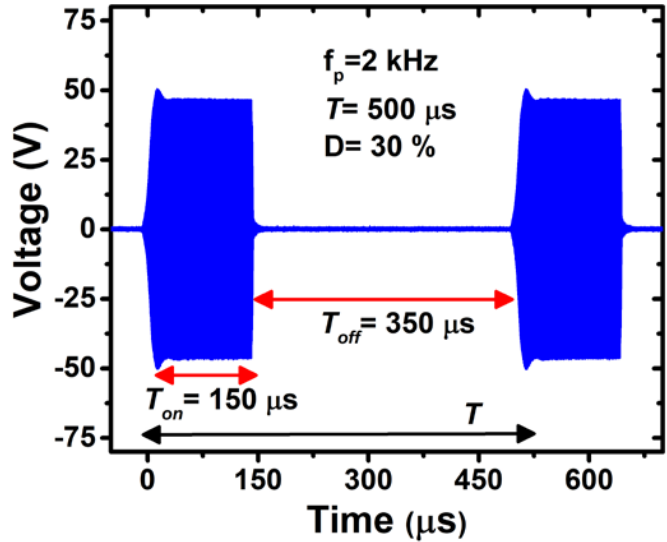

**Figure 4:** Typical waveform of pulse-modulated RF voltage for 2 kHz, 30% duty cycle, 50W power, and 1.5 lpm gas flow rate.

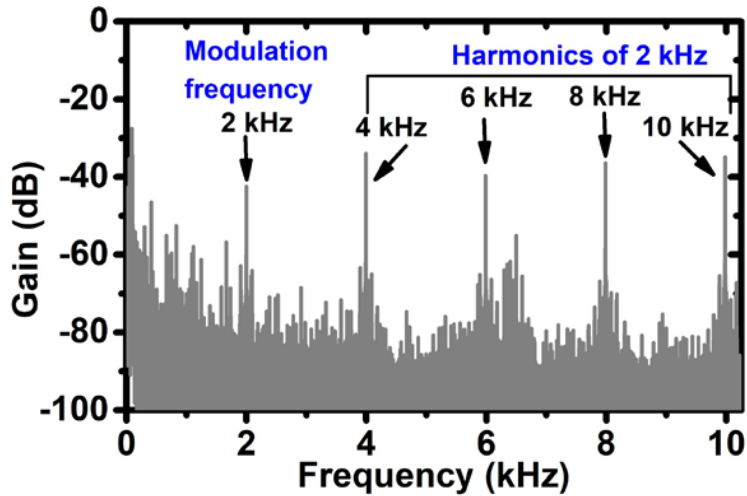

**Figure 5:** Frequency spectrum of the measured acoustic signal showing the fundamental and its harmonics for 2 kHz, 30% duty cycle, 50 W power, and 1.5 lpm gas flow rate. The acoustic signal is recorded by arranging the microphone near the nozzle by keeping a 5 mm distance radially to the plasma jet.

It is possible that the pulsing of this pressure gradient could also produce electrostatic acoustic plasma waves carrying orbital angular momentum (OAM). [44–49] In the present case, the generation of such waves could be understood by following explanation. The frequency of such acoustic waves would correspond to the fundamental pulsing frequency (= 2 kHz in present case) and its harmonics (shown in Figure 5). Waves carrying orbital angular momentum have been predicted in the literature for long, both for light beams as well as for electrostatic electron and acoustic plasma waves in plasmas.[44–49] *A major property of such waves is that these have helical constant phase surfaces as well as a propagation vector K that traces a helical path in space.*[29,46] Since the wave is electrostatic, its electric field would have to be aligned along $\vec{K}$ and *confined within the plasma filament*. Hence, the plasma filament along with glow that is formed under these circumstances also follows the helical trajectory of the electric field and $\vec{K}$. The real part of the $\vec{K}$ has an azimuthal part and an axial part and can be expressed as,

$$\vec{K} = (L/r_p)\hat{e}_\theta + K_z\hat{e}_z$$

where $L$ is the azimuthal mode number (OAM topological mode number) and is proportional to the orbital angular momentum[45,46,48], $K_z$ is the axial propagation constant, $r_p$ is the radius of the filament contour, $\hat{e}_\theta$ and $\hat{e}_z$ are the unit vectors along the azimuthal direction, $\theta$ and the axial direction, $z$, respectively. An illustration of the single-mode helical wave with a value $L = +1$ is shown in Figure 6. The momentum carried by a *quantum of the acoustic wave plasmon* in the filament is, $\vec{P} = \hbar \vec{K}$ where $\hbar$ is reduced Planck's constant ($= h/2\pi$) *and* the angular momentum carried by the plasmon is $\vec{M} = \vec{R} \times \vec{P}$ where $\vec{R}$ is the radial position of the plasmon given as, $\vec{R} = r_p\hat{e}_r$. Evaluating gives, $\vec{M} = \hbar [L\,\hat{e}_z - r_p K_z \hat{e}_\theta]$. Thus the orbital angular momentum along the axial direction is $\hbar L$, while that along with the azimuthal direction averages to zero as the plasmon propagates along the helical path of filament.

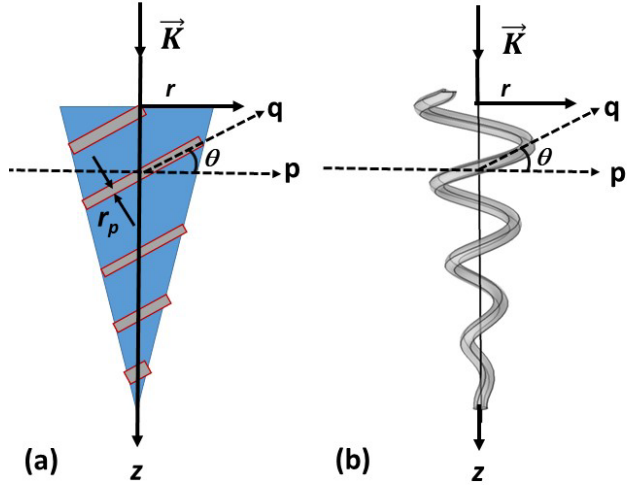

**Figure 6:** Pictorial representation of propagating helical plasma jet with right handed (clockwise) OAM. (a) 2-dimensional representation (b) 3-dimensional representation.

It is possible therefore, that the formation and existence of the helical plasma filament could be quite intimately linked to that of acoustic waves with orbital angular momentum, which in turn can arise from the pulsating plasma gradients. However, detailed simulation studies and dedicated experiments need to be developed for greater understanding of the discharge dynamics of this kind of plasma shape. On the other hand, in CW mode, there is almost uniform gas temperature distribution, and since a pulsating pressure gradient is not available, the periodic time dependence required to excite the acoustic waves and hence the helical plasma filament is also missing. Therefore, the flow is smooth and follows classical diffusion-dominated transport.

## D. Influence of input parameters

In our experimental study, it was found that the input parameters viz., duty cycle, gas flow rate, and pulsed RF power significantly influence the formation of the helical structure and their geometrical features which are discussed in the following section.

### i) Effect of duty cycle

Figure 7 shows the effect of the duty cycle on the evolution of helical structures at 50 W pulse-modulated RF power, and 2 kHz pulsing frequency. A noteworthy response is observed when the duty cycle is increased from 10% to 80%. Corresponding average power ($P_{avg}$), $T_{on}$, $T_{off}$ and corresponding average gas temperature values are shown in Table 1. Helical structures become more pronounced for duty cycles more than 10%, because the periodic pressure difference generated due to the gas heating and cooling produces sufficient impact to form the structures. At higher duty cycles (from 50-80%) though the relative intensity of the plasma increases, the *helical shape disappeared monotonically and plasma acquired quasi-continuous wave like shape, leading to a regular conical shape*. This is to say, at higher duty cycles $T_{off}$ is small compared to $T_{on}$, therefore the time interval for gas cooling becomes lower and the influence of pressure pulsation becomes less significant as compared to that at lower duty cycles. For example, at 80 % duty cycle, the average gas temperature (~ 425 K) is high due to continuous heating at a high average power (i.e. 40 W) over a long on-duration of ~ 400 µs, which is four times the plasma cooling time (i.e. 100 µs). Thus the gas does not get enough time to cool significantly in the off duration and the temperature remains almost constant during on and off durations. As a result, pressure pulsations within the discharge region become weaker and the plasma tends to retain a reasonably conical shape. Whereas at 30 % duty cycle, the average gas temperature (~ 339 K) is comparatively low because the plasma is on only for 150 µs and remains off for rest of the period (i.e. 350 µs). Therefore, cooling is faster and the discharge region and consequently, the pressure pulsation experiences significant gas temperature variation in one modulation cycle, which leads to the formation of the helical structure. Thus, the duty cycle in pulse operation emerges as one of the control parameters for the generation of the helical structure, especially for the case when $T_{off} \geq T_{on}$.

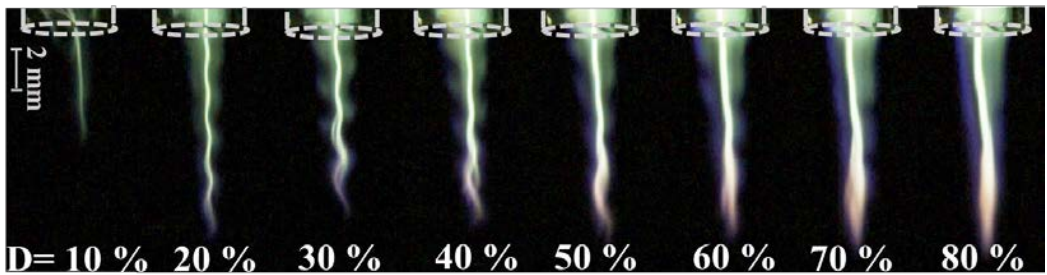

**Figure 7:** Effect of pulse duty cycle on plasma jet shape (front view) at 50 W, 1.5 lpm, and 2 kHz, pulse frequency. Increasing duty cycle increases gas temperature, intensifies the plasma jet and suppresses the helical structures.

**Table1:** Average power, pulse duration and gas temperature ($T_g$) with reference to the duty cycle for 50 W pulse-modulated RF power, 2 kHz pulsing frequency and 1.5 lpm gas flow rate.

| Duty cycle D (%) | Average applied power $P_{avg}$ (W) | Pulse on duration $T_{on}$ (µs) | Pulse off duration $T_{off}$ (µs) | Average gas temperature $T_g \pm 2$ K |
|---|---|---|---|---|
| 10 | 5  | 50  | 450 | 312 |
| 20 | 10 | 100 | 400 | 325 |
| 30 | 15 | 150 | 350 | 339 |
| 40 | 20 | 200 | 300 | 356 |
| 50 | 25 | 250 | 250 | 371 |
| 60 | 30 | 300 | 200 | 392 |
| 70 | 35 | 350 | 150 | 406 |
| 80 | 40 | 400 | 100 | 425 |

The variation of geometrical features such as plasma jet length and width at the nozzle with the variation in duty cycle is given in Figure 8. Increment of duty cycle from 10% to 20% sharply increases the length from 5.5 mm to 11 mm. A slight dip in the length is noticed at 30% duty cycle and thereafter it increases with increment in duty cycle. This dip in the length could be related to the enhancement of helical shape. Thereafter increment could be associated with an increased average power supply with the duty cycle (i.e., applied pulse duration). The plasma jet width at the nozzle shown in Figure 8, increases with duty cycle and saturates at ~ 2.75 mm beyond 50% duty cycle. The increment in the plasma width at the nozzle is related to the increment in applied power. For higher duty cycle, plasma does not spread radially (even when average power is increased), since the plasma begins to fill the interior of the pyrex tube for a fixed gas flow rate.

Experiments were also performed to understand the influence of *fixed average power* (~ 15 W) *with increase in duty cycle* (*D*) from ~ 10 to 80%. It was seen that at low duty cycles (*D* = 10 – 20 %) the jet is long (~ 2 cm) with helical structure beginning to form. At *D* ~ 30 – 50 %, the helical structure becomes more prominent, though there is a shortening of the length. With further increase in *D*, the length of the jet decreases progressively, together with a smoothening of the helical shape with the minimum jet length being seen for *D* = 80 %. In general, the *progressive disappearance of the helical shape with increasing D vibes with the fact that to produce an adequate helical shape of the jet, a sufficient cooling time needs to be provided,* i.e., $T_{off} > T_{on}$.

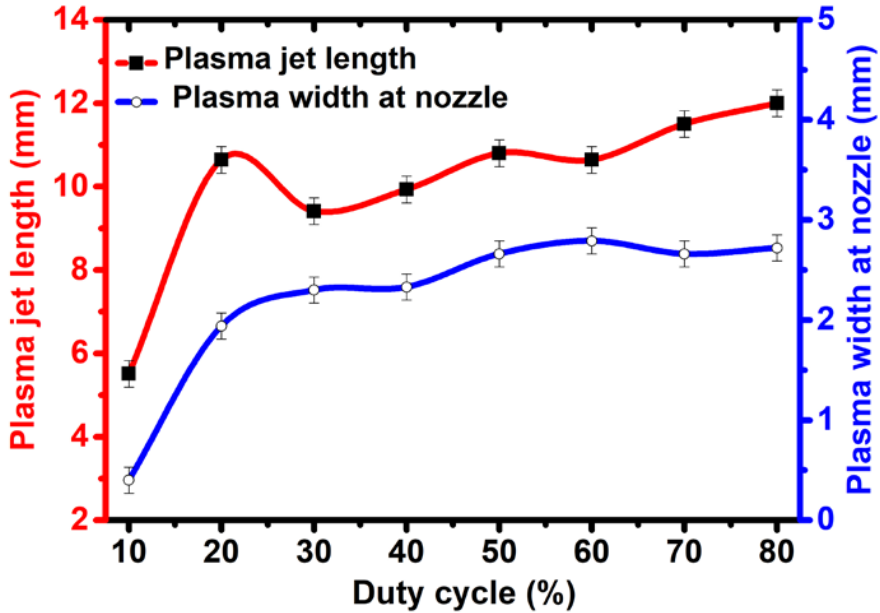

**Figure 8:** Effect of duty cycle on plasma jet length and plasma width at nozzle at 2 kHz PM, 50 W power, 1.5 lpm gas flow

### ii) Effect of gas flow rate (Q)

To understand the effect of gas flow rate on the dynamics of the helical shape of plasma jet (i.e., formation, expansion, and dissolving) the images are captured by arranging the camera, (a) at an angle of 45° to the discharge axis facing the discharge (argon gas flow rate varied from 0.5 - 2.5 lpm) and (b) to capture the front view, perpendicular to the axial direction (flow rate varied from 1.5 - 12 lpm).

As shown in Figure 9 (a), at 0.5 lpm, the short filament generated is inclined to the inner surface of the glass tube. With a gradual increase in the gas flow till 1 lpm, the filament surrounded by the glow discharge continues to increase in length and maintains its attachment to the inner surface of the glass tube. Further increment in gas flow aligns the filament along the reference axis, but with initiation of a little helical-like structure. As the gas flow rate is increased from 1.2-1.5 lpm, the helical patterns become more prominent and the clockwise turns appear in the ambient air. The corresponding Reynolds number of neutral argon gas for the given geometry is $R_e$ = 837 for 1.5 lpm, which shows that gas flow is in the laminar regime. The gas flow remains in the laminar region up to 3 lpm after which the flow enters into the transition mode and post 7.5 lpm, the gas flow becomes turbulent.[50] Images in Figure 9 (a) further strengthen the observed plasma jet has a helical pattern with reference to its axis. From 2 lpm gas flow rate, the helical structures are disturbed with reduced jet length. This disturbance of the helical structures occurs due to the additional velocity increment by plasma onset which leads to the onset of turbulence earlier, even though the neutral flow is still in the laminar regime similar to earlier reports.[15,21]

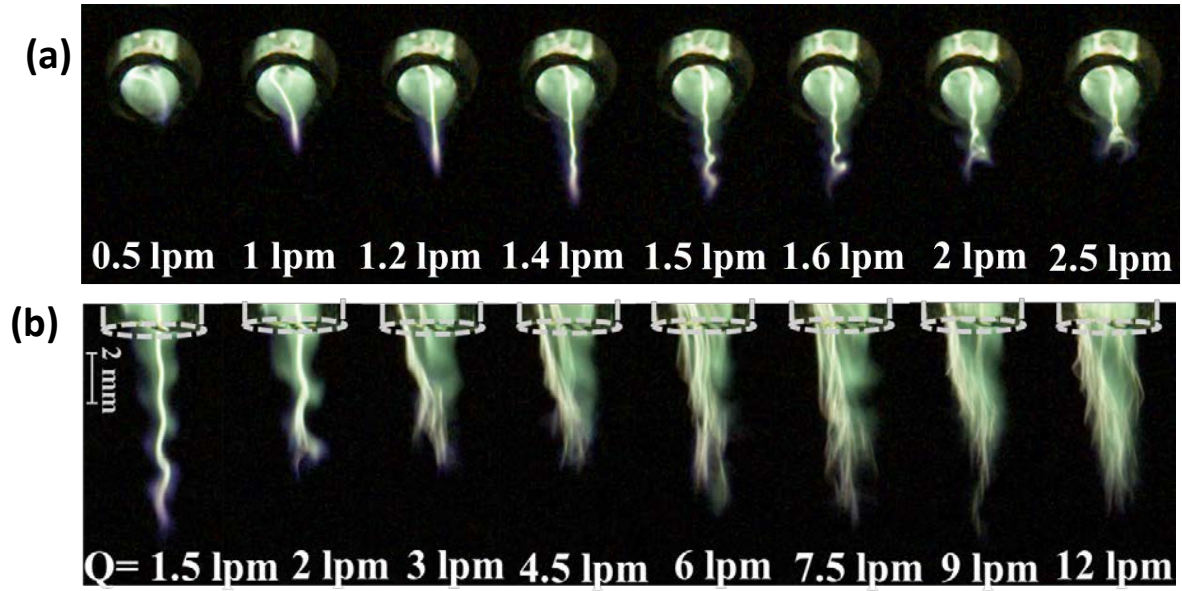

**Figure 9:** Images of helical plasma jet as a function of the argon gas flow rate. (a) Images are captured at a camera angle of 45° from -$z$-axis (tilted view) with (0.5 - 2.5 lpm) gas flow rate depicts the helical structure formation (b) gas flow rate 1.5-12 lpm and images are captured perpendicular to the axial direction (front view) shows the dynamics. Higher gas flow rate leads to disturbed structure, reduced plasma jet length, and radial spread of plasma. All the images are captured with fixed 50 W power, 2 kHz, 30% duty cycle

The dynamics of the helical plasma jet with increasing argon gas flow rate in the range of 1.5-12 lpm is shown in Figure 9(b). The images show the length of the plasma jet is maximum (10 mm) at 1.5 lpm which lies in the laminar mode of gas flow and has an impeccable helical structure. With an increase in gas flow to 2 lpm, the shape of the plasma jet starts to disrupt and the number of turns and the overall length of the plasma jet reduces. This is attributed to the early onset of turbulence [23,24,51–53] disturbing the normal flow of plasma in the ambient air. A further increment in the gas flow rate (up to 4.5 lpm) completely disturbs the plasma pattern and a reduction in length and a spreading of the radial width of the plasma jet is noticed. At higher flow rates (from 6 lpm-12 lpm) the gas channel becomes completely turbulent. In this region, surprisingly, the plasma length increases, which could be due to increased ionization.[50]

The plasma jet length and the corresponding Reynolds number as a function of gas flow rate is given in Figure 10, which shows the maximum length of plasma jet is attained at 1.5 lpm where the impeccable helical patterns also formed. Thus, an optimized gas flow is required to form and propagate pulse-induced helical structures. In the present case, the helical shape of plasma is observed only at a low gas flow rates, within a narrow range.

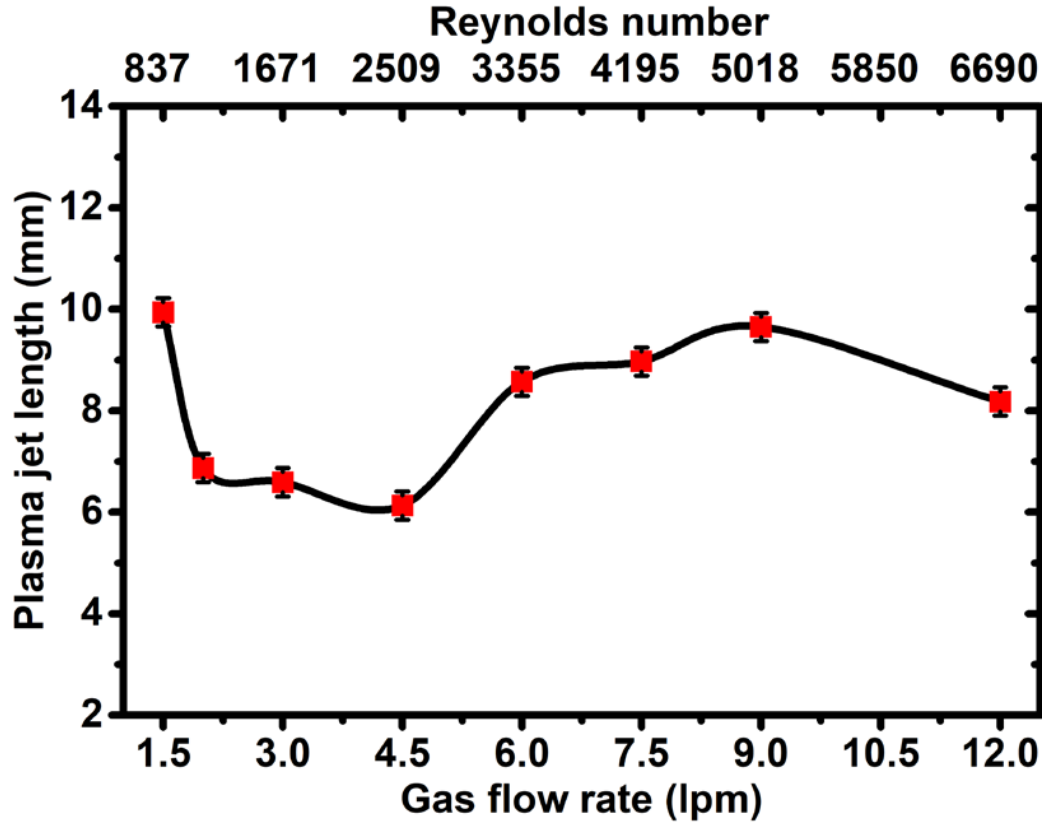

**Figure 10:** Plasma jet length as a function of argon gas flow rate at 50W applied power, 2 kHz, 30% duty cycle.

### iii. Effect of applied power

The evolution of the helical shape with respect to applied pulsed RF power is investigated by varying the power from 20- 50 W for 30% duty cycle as shown in Figure 11. It is interesting to observe that the number of turns of the plasma jet increases up to 30 W and remains constant up to 50 W. The increment in the number of turns and plasma length are associated with applied power; although the number of turns increases, the pitch ($\lambda$) of the helical structure remains practically the same. This behavior is due to generation of *sufficient gas temperature pulsation between the on and off duration* of the plasma at high applied power. For example, at 50 W RF power, average gas temperature of the plasma is adequately high (~ 339 K), from which one may infer that in the 'on' duration of the pulse, this temperature is able to drop almost completely during the longer off duration of the cycle. In effect, an ample amount of time in the pulse off duration allows sufficient lowering of the temperature within the discharge region, which helps to generate a pronounced pressure pulsation that leads to the formation of significant helical structures. Contrarily, at low applied powers (say 20 W) the associated gas temperatures are comparatively low (~ 315 K) and the discharge region is not heated much during the 'on' duration of the pulse. As a result, the temperature pulsations during the modulation cycle are not sufficiently strong and only a weak pressure gradient is generated so that the helical structures produced are also less prominent. More gas heating at higher power with longer cooling period produces greater pressure pulsation and therefore, more pronounced and sharper helical structures are generated. Further, the variation of plasma length and width at the nozzle as a function of power is shown in Figure 12, which shows the plasma length and width at nozzle increase up to 30W, after which it tends to remain almost constant for higher powers. In the present experiments, power was restricted to 50 W due to the mode transition from glow to arc.

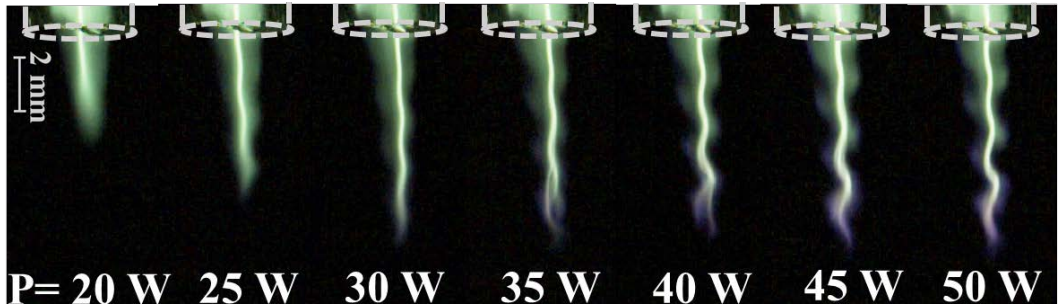

**Figure. 11** Effect of RF power on the helical shape of plasma jet for 2 kHz, 1.5 lpm, and 30% duty cycle.

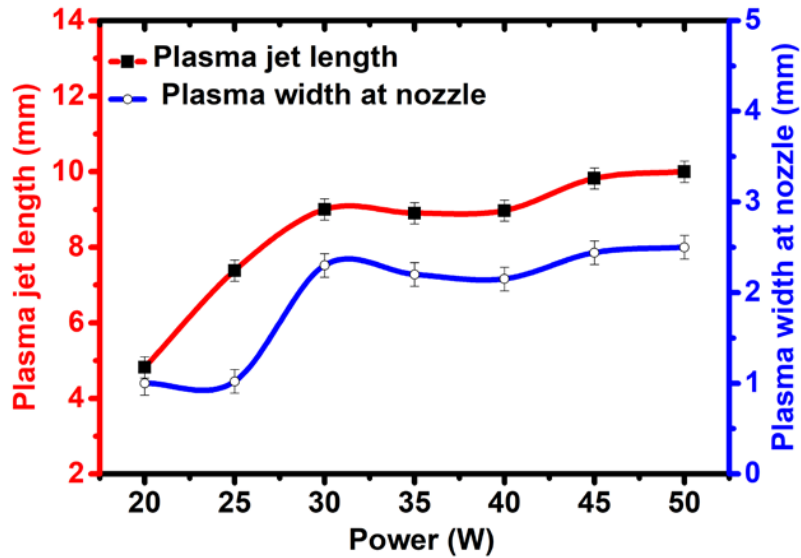

**Figure 12:** Variation of plasma jet length and plasma width at the nozzle as a function of applied power for fixed gas flow rate 1.5 lpm, 2 kHz, 30% duty cycle.

It is worth noting that the viscosity of a fluid is an important property that decides the behavior of the flow. It is well known that the viscosity of a neutral gas increases with the temperature. As mentioned earlier, the *average* gas temperature increases with increase in the duty cycle, which is associated with higher average power applied to the plasma. It was seen that the shorter times available for gas cooling at higher duty cycles cause the suppression of helical structures. It must be stated that there is also the possibility that a high viscosity could be associated to higher duty cycles that could be responsible for the diminishing of the helical structure. An increment in the Ar viscosity with gas temperature opposes the pressure pulsing forces which further stabilizes the plasma flow[52] and the plasma jet remains smooth and conical. However, in the case of power variation, the maximum applied RF power in this work is 50 W with a 30 % duty cycle that corresponds to the average power of 15 W only. Associated average gas temperature in the present case (~ 339 K for 15 W average applied power) is very low as compared to the maximum duty cycle case (~ 425 K for 40 W average applied power). Therefore, the viscosity effect due to the gas temperature is insignificant for low average power, and the plasma can acquire a helical shape.

## IV. CONCLUSIONS

In conclusion, this paper reports a helical argon plasma jet, excited with 2 kHz pulse modulated RF power. This helical shape of the plasma jet is very different from the conventional conical plasma jets. It is found that helical structures in the plasma jet can be generated if optimum temperature and pressure difference are maintained. The selection of an appropriate duty cycle with $T_{off} \geq T_{on}$ for the pulse modulated RF power, along with appropriate gas flow rates, play a significant role in the formation of these helical structures. Probably the acoustic wave generated orbital angular momentum plays a role in developing the helical structures. However, a deeper understanding of such a phenomenon demands further experimental and simulation studies. Such helical patterns are of importance since they facilitate more air entrainment into the plasma ionization region, enhancing formation of reactive oxygen and nitrogen species (RONS) at a lower temperature, which renders them important for biomedical applications.

## DATA AVAILABILITY
The data that support the findings of this study are available from the corresponding author upon reasonable request.